\documentclass[proof]{pasj00}

\usepackage{times}
\usepackage{ulem}

\begin{document}

\SetRunningHead{Nakajima et al.}{A 0535+26 Outbursts in 2009-2011 by MAXI and Swift}


\title{Precursors and Outbursts of A 0535+26 in 2009-2011 \\
observed by the MAXI/GSC and the Swift/BAT}
\author{
        Motoki \textsc{Nakajima}, \altaffilmark{1}
        Tatehiro \textsc{Mihara}, \altaffilmark{2}
        Mutsumi \textsc{Sugizaki}, \altaffilmark{2}
        Motoko \textsc{Serino}, \altaffilmark{2}
  	Masaru \textsc{Matsuoka}, \altaffilmark{2}
        \\
        Nobuyuki \textsc{Kawai}, \altaffilmark{3}
        Kazuo \textsc{Makishima}, \altaffilmark{2,4,5}
}
\altaffiltext{1}{School of Dentistry at Matsudo, Nihon University, 
  2-870-1, Sakaecho-nishi, Matsudo, Chiba, JAPAN 271-8587}
\altaffiltext{2}{MAXI team, RIKEN, 
  2-1 Hirosawa, Wako, Saitama, JAPAN 351-0198}
\altaffiltext{3}{Department of Physics, Tokyo Institute of Technology, 
  2-12-1 Ookayama, Meguro-ku, Tokyo, JAPAN 152-8551}
\altaffiltext{4}{Department of Physics, University of Tokyo, 
  7-3-1 Hongo, Bunkyo-ku, Tokyo, JAPAN 113-0033}
\altaffiltext{5}{Research Center for the Early Universe, University of Tokyo, 
  7-3-1 Hongo, Bunkyo-ku, Tokyo, JAPAN 113-0033}

\email{(MN) nakajima.motoki@nihon-u.ac.jp}


\maketitle

\begin{abstract}

  Over the 3-year active period from 2008 September to 2011 November,
  the outburst behavior of the Be/X-ray binary A 0535+26 was
  continuously monitored with the MAXI/GSC and the Swift/BAT.  The
  source exhibited nine outbursts, every binary revolution of 111.1
  days, of which two are categorized into the giant (type-II)
  outbursts.  The recurrence period of these outbursts is found to be
  $\sim115$ days, significantly longer than the orbital period of
  111.1 days.  With the MAXI/GSC, a low-level active period, or a
  ``precursor'', was detected prior to at least four giant outbursts.
  The precursor recurrence period agrees with that of the giant
  outbursts.  The period difference of the giant outbursts from the
  orbital period is possibly related with some structures in the
  circumstellar disc formed around the Be companion. Two scenarios,
  one based on a one-armed disc structure and the other a Be-disc
  precession, are discussed.

\end{abstract}


\section{Introduction}
\label{sec1}

Be/X-ray binaries are a group of high-mass X-ray binaries that consist
of a neutron star and an OB-type star with Balmer emission lines (Be
star). They are major members of recurrent X-ray transients and are
usually detected when they exhibit bright X-ray outbursts lasting for
a week to months.  During the outbursts, the X-ray luminosities
increase by 3--5 orders of magnitude from those in the quiescent
states.  Although the mechanism of these outbursts has not been fully
understood, it is considered to be related with mass accretion from
the circumstellar disc of the Be star onto the neutron star.

X-ray outbursts of Be/X-ray binaries are classified into two types,
normal outbursts (type-I) and giant ones (type-II)
(e.g. \cite{reig2011}).  The former usually occur when the neutron
star passes through the periastron, and hence repeat by the orbital
period.  Their peak luminosities distribute in a narrow range of
$10^{36} - 10^{37}$ erg s$^{-1}$.  Some of those sources showed
moderate spin-up of the neutron star during the normal outbursts
(e.g. GS~1843$-$02; \cite{finger1999}).  On the other hand, the
latter, namely giant outbursts, are characterised by their higher
X-ray luminosity (typically $\geq10^{37}$ erg s$^{-1}$), longer
duration, less frequent occurrence, and their orbital phases out of
the periastron \citep{priedhorsky1983,reig2011}.  Since large spin-up
episodes of the neutron star often accompany giant outbursts, an
accretion disc is considered to form around the neutron star in
occasions \citep{reig2011}.  In addition to the observational outburst
classification, a new scenario based on the analytical/numerical
studies \citep{okazaki2013} is proposed to explain the two type of
outbursts by different mass accretion mechanism.

A 0535+26 is one of the representative Be/X-ray binaries in our
Galaxy.  The source was discovered by the Ariel-5 satellite
\citep{rosenberg1975,coe1975} when it exhibited a giant outburst in
1975.  The subsequent observations revealed its 103-s coherent X-ray
pulsation \citep{rosenberg1975}.  The optical counterpart is the O9.7
IIIe star HD245770 at a distance of 1.8$\pm$0.6 kpc
\citep{giangrande1980}.  \citet{steele1998} revised the spectral type
of its optical counterpart with B0 IIIe.  Its orbital elements,
including the orbital period (111.1$\pm$0.1 day), eccentricity
(0.47$\pm$0.02) and the epoch of the periastron passage (MJD=53613.0),
were determined by \citet{finger2006}.  Both normal and giant
outbursts have been observed so far \citep{motch1991}.  The giant
outbursts were observed nine times, typically every a few years
(\cite{camero2012} and references therein).  All these events occurred
after the periastron passages, and were followed by a sequence of
normal outbursts.

In 2008 September, the {\it RXTE}/ASM detected a renewed activity of A
0535+26, since the previous active period which ended in 2005
\citep{levine2008}.  The new active period continued three years until
2011 November, and the source meanwhile exhibited nine outbursts
including both the normal and the giant ones.  This is the longest
active period that has ever been observed since its discovery
\citep{finger1996,camero2012}.

In this paper, we report on the X-ray outbursts of A 0535+26 during
the above active period, based on continuous monitoring by the
MAXI/GSC \citep{matsuoka2009} and the Swift/BAT \citep{gehrels2004}.
The GSC onboard MAXI successfully detected not only normal / giant
outbursts but also low-intensity activity during the outburst
intermission \citep{sugizaki2009}.

\section{Observations}
\label{sec2}

\begin{figure*}
  \begin{center}
    \FigureFile(160mm,){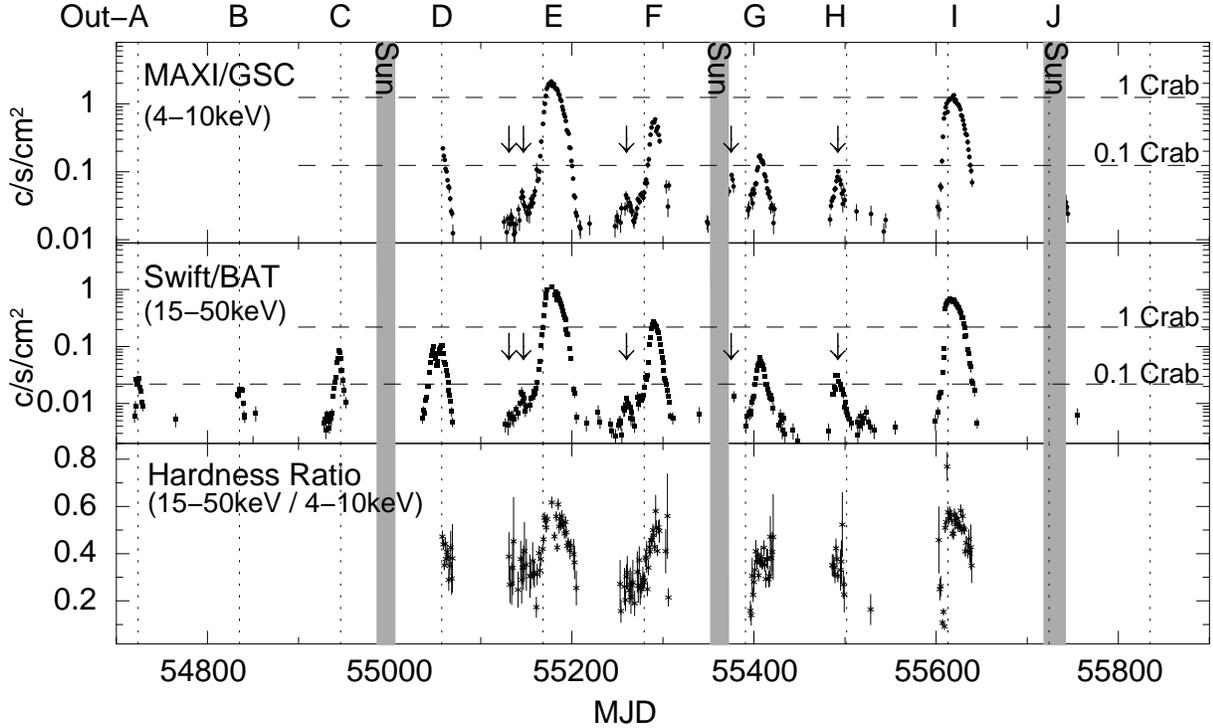}
  \end{center}
  \caption{ Light curves of A 0535+26 from 2008 September to 2011
    November, in the 4--10 keV band by the MAXI/GSC (top : 1 Crab $=
    1.24$ c cm$^{-2}$ s$^{-1}$) and in the 15--50 keV band by the
    Swift/BAT (Middle : 1 Crab $= 0.22$ c cm$^{-2}$ s$^{-1}$).  The
    horizontal dashed lines indicate the flux level in Crab units.
    The bottom panel shows hardness ratio of the Swift/BAT to the
    MAXI/GSC.  The vertical dotted lines, the arrows, and the gray
    belts represent the periastron passages \citep{finger2006},
    precursors, and periods of the Sun avoidance of the MAXI/GSC,
    respectively. }
\label{f1}
\end{figure*}

\subsection{MAXI/GSC}
\label{sec2.1}

Since the in-orbit operation started in 2009 August, the Gas Slit
Camera (GSC) \citep{mihara2011, sugizaki2011} onboard MAXI on the
International Space Station scans almost the entire sky in every ISS
orbital period of about 92 minutes.  For $\sim45$ s typically every 92
minutes, the GSC observes an X-ray source in any position on the sky,
except an area which is within $\lesssim 10^\circ$ of the Sun.  In the
web page\footnote{http://maxi.riken.jp/top/}, the MAXI team provides
GSC light curves of about 300 cataloged X-ray sources including A
0535+26, for three energy bands of 2--4, 4--10 and 10--20 keV.  Since
the data in the 4--10 keV band have the highest signal-to-noise ratio,
we hereafter use this energy band to extract the MAXI/GSC light
curves.

\subsection{Swift/BAT}
\label{sec2.2}

The Swift Burst Alert Telescope (BAT; \cite{barthelmy2005}) has
performed all-sky hard X-ray survey in the 15--50 keV energy band
since the operation started in 2004.  The Swift/BAT team provides
light curves for about a thousand of cataloged X/gamma-ray sources via
the archive web
page\footnote{http://swift.gsfc.nasa.gov/docs/swift/results/transients/}.
We obtained the light curve data of A 0535+26, and utilized
good-quality data selected with a quality flag $=0$.

\section{Analysis and Results}
\label{sec3}

\subsection{Long-term Light Curve and Hardness Ratio}
\label{sec3.1}

Figure \ref{f1} shows the X-ray light curves of A 0535+26 observed
with the MAXI/GSC (2009 August 15 to 2011 November 30) and the
Swift/BAT (2008 September 1 to 2011 November 30), over those periods
when significant flux was detected above 3 $\sigma$ of the statistical
uncertainty.  The hardness ratio of the Swift/BAT band to the MAXI/GSC
band is shown in the bottom panel.  Nine sequential outbursts are
clearly recognized in the Swift/BAT light curve with 111.1-days
orbital cycle.  As shown in figure \ref{f1}, we hereafter call them
Out-A, B, C, ..., and J.

The first three outbursts, namely Out-A, B, and C, exhibited their
peaks at times approximately consistent with periastron passages.  The
peak intensity of Out-A and Out-B are all $\sim 0.1$ Crab in the
15--50 keV band, while that of Out-C is about 4 times higher, $\sim
0.4$ Crab.  The outburst profile changed clearly after Out-D: the
profiles of Out-A, B, and C are single-peaked, whereas Out-D has a
double-peaked profile \citep{caballero2011,caballero2013}.

After Out-D, MAXI started the operation in orbit, and as indicated by
arrows in figure \ref{f1}, the GSC started detecting even periods of
weaker X-ray emission at an earlier orbital phase than the periastron
\citep{sugizaki2009}.  They are more easily recognized in the zoomed
light curves in figure \ref{f2}.  We call this type of event a
``precursor''.  These precursors were also detected with the BAT
(figure \ref{f1} middle).  In particular, those preceding Out-E are
found to involve little changes in the hardness ratio (figure \ref{f1}
bottom).

The subsequent outburst, Out-E, developed into a giant one, and the
peak 4--10 keV intensity reached $\sim 1.8$ Crab.  The source reached
the outburst maximum 9.3 d after the periastron, and the hardness
ratio changed significantly across the outburst.  Assuming that the
X-ray emission is isotropic and the spectrum shape is like the Crab
nebula, the total X-ray energy released during the outburst period
(MJD=55157--55201) in the 4--10 keV band is estimated to be $1.1
\times 10^{43}$ erg from the daily MAXI/GSC light curve.

The next outburst, Out-F, also developed into a large one, and the
peak phase was further delayed.  A precursor was also seen $\sim 30$
days before Out-F.  The next precursor preceding Out-G was observed by
the MAXI/GSC but not by the Swift/BAT due to the Sun angle limit.
Fermi/GBM pulsar monitor detected an increase of pulsed flux both in
the precursor and the mainbody of Out-G \citep{camero2012}.  The peak
of Out-G was also delayed from the periastron.  The hardness ratio
also increased similarly as the outburst developed.  However, it did
not reach the level of the peak of Out-E.  The change of the peak
intensity apparently has a positive correlation with the peak hardness
in these outbursts.

In Out-H, the intensity reached a maximum before the periastron.  No
activity was seen before the peak in the MAXI/GSC or the Swift/BAT
data.  Instead, a possible post-outburst activity is seen in the
Swift/BAT data on MJD$\sim$55523, or at on orbital phase of $\sim0.2$.

After the relatively weak Out-H, we observed yet another giant
outburst, Out-I, which exhibited the peak 4-10 keV intensity of
$\sim1$ Crab.  No precursor was seen in either the MAXI/GSC or the
Swift/BAT data.  The hardness ratio changed with the flux similarly as
in Out-E.

The next periastron after Out-I was observable with neither mission
due to the Sun avoidance.  However, the positive flux detected on
MJD$\sim$55745 suggests the presence of another outburst (Out-J).
This possibly faint outburst was not observed by the Fermi/GBM.  After
that, no significant flux have been observed from the source till the
end of 2012.

In addition to the X-ray activities, significant optical variations
have also been reported in these period
\citep{moritani2011,camero2012,yan2012}. The V-band magnitude started
to decrease in the epoch of Out-A and then reach the quiescent level
in the Out-D. On the other hand, the equivalent width of H$\alpha$
line increased around the Out-D and then decreased after the Out-F.

\subsection{Evolution of Outburst Orbital Phase}
\label{sec3.2}

\begin{figure}
  \begin{center}
  \FigureFile(85mm,){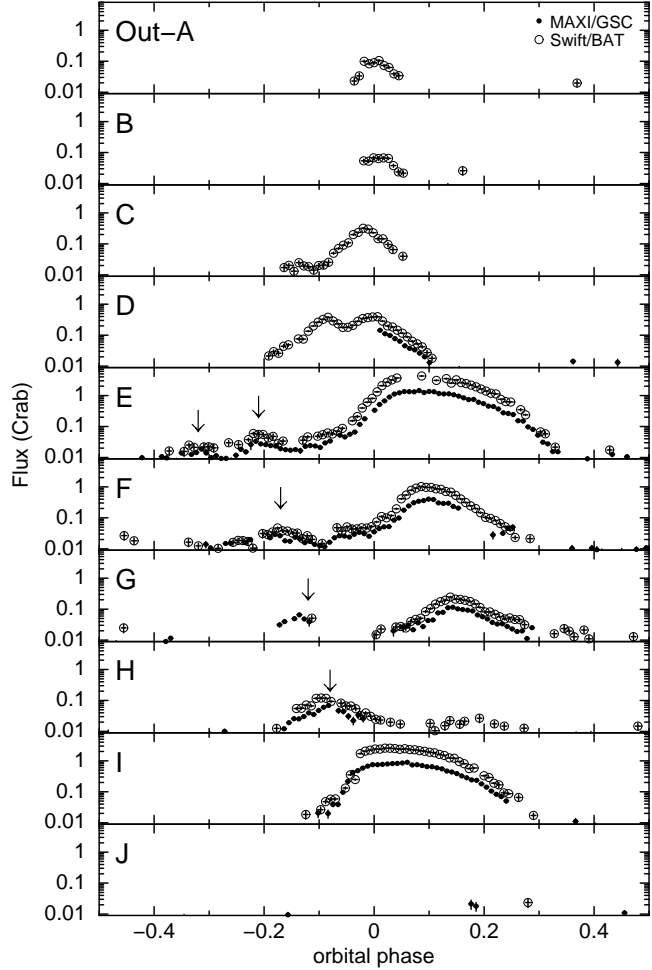}
  \end{center}
  \caption{ The same light curves as in figure \ref{f1}, but shown for
    individual 111.1-d orbital cycles and those fluxes in Crab units.
    The MAXI/GSC(4--10 keV) and the Swift/BAT(15--50 keV) data are
    plotted in the same panel with different marks.  The flux
    conversion factors between the count rates and Crab units are same
    as in figure \ref{f1}. The origin of the orbital phase ($=0$)
    corresponds to the periastron passage.  Five arrows in panels E,
    F, G, and H indicate precursor peaks. }
\label{f2}
\end{figure}

As described in the previous section, the outbursts (Out-E, F and G)
and the precursors both exhibited significant offsets in orbital phase
from the periastron passages.  In this paper, we calculated the epoch
of periastron passages using the orbital elements in
\citet{finger2006}, which were determined from X-ray observations in
2005.  Judging from the precision of the orbital period (111.1$\pm$0.1
d), the error of the extrapolated periastron time in 2009--2011 is
$\sim$1 d at most.  In order to better visualize this effect, figure
\ref{f2} shows the same light curves, but in cycle-by-cycle form,
against the orbital phase.  Thus, we find a clear systematic shift,
from Out-E to Out-G, in the orbital phase of the outburst peak.  The
precursor phase (marked by arrows in figure \ref{f2}) also shifted
similarly to the main peak.  Among these outbursts, the intervals
between the precursor and the outburst are almost constant at $\sim
30$ days (or $\sim 0.27$ orbital phase).

Interestingly, Out-H, which did not develop into a large one,
consisted of two peaks like Out-D.  The former event can be identified
as a precursor based on an orbital-phase extrapolation from Out-E, F,
and G.  The second peak can be considered as a main outburst, although
its intensity is smaller than that of typical outburst.  In the next
outburst, Out-I, the peak just came after the periastron, and neither
a precursor nor a post-outburst flare were observed.

\begin{table*}[tb]
  \caption{Dates and orbital phases of outbursts and precursors}
  \label{t1}
  \begin{center}
    \begin{tabular}{rlcccccc}\hline
      & & & \multicolumn{2}{c}{MAXI} & & \multicolumn{2}{c}{Swift} 
      \\ \cline{4-5} \cline{7-8} 
      & & & MJD & orbital phase$^a$ & & MJD & orbital phase$^a$ \\ \hline 
      outburst & A & & ----- & -----
                 & & 54,724.5 $\pm$ 0.2 & 0.0045 $\pm$ 0.0009 \\
       & B & & ----- & -----
                 & & 54,835.9 $\pm$ 0.5 & 0.0072 $\pm$ 0.0045 \\
       & C & & ----- & -----
                 & & 54,944.5 $\pm$ 0.1 & 0.9847 $\pm$ 0.0009 \\
       & D & & ----- & ----- 
                 & & 55,047.1 $\pm$ 0.1 & 0.9082 $\pm$ 0.0009 \\
       & D$^b$ & & ----- & ----- 
                 & & 55,056.9 $\pm$ 0.1 & 0.9964 $\pm$ 0.0009 \\
       & E & & 55,177.7 $\pm$ 0.2 & 0.0837 $\pm$ 0.0018 
                 & & 55,178.4 $\pm$ 0.2 & 0.0900 $\pm$ 0.0018 \\
       & F & & 55,291.1 $\pm$ 0.1 & 0.1044 $\pm$ 0.0009 
                 & & 55,289.7 $\pm$ 0.3 & 0.0918 $\pm$ 0.0027 \\
       & G & & 55,407.3 $\pm$ 0.1 & 0.1503 $\pm$ 0.0009 
                 & & 55,407.4 $\pm$ 0.2 & 0.1512 $\pm$ 0.0018 \\
       & H & & ----- & ----- 
                 & & 55,521.9 $\pm$ 3.0 & 0.1818 $\pm$ 0.0270 \\
       & I & & 55,617.9 $\pm$ 0.1 & 0.0459 $\pm$ 0.0009 
                 & & 55,617.4 $\pm$ 0.2 & 0.0414 $\pm$ 0.0018 \\
      \hline
      precursor & E & & 55,132.4 $\pm$ 1.9 & 0.6760 $\pm$ 0.0171 
                  & & 55,132.9 $\pm$ 0.7 & 0.6805 $\pm$ 0.0063 \\
       & E$^c$ & & 55,146.2 $\pm$ 0.8 & 0.8002 $\pm$ 0.0072
                      & & 55,145.7 $\pm$ 0.7 & 0.7957 $\pm$ 0.0063 \\
       & F & & 55,259.8 $\pm$ 0.5 & 0.8227 $\pm$ 0.0045
                  & & 55,261.2 $\pm $0.7 & 0.8353 $\pm$ 0.0063 \\
       & G & & 55,378.4 $\pm$ 1.1 & 0.8902 $\pm$ 0.0099 
                  & & ----- & ----- \\
       & H & & 55,492.5 $\pm$ 0.3 & 0.9172 $\pm$ 0.0027 
                  & & 55.492.5 $\pm$ 0.2 & 0.9172 $\pm$ 0.0018 \\
      \hline
      \multicolumn{5}{l}{$^a$:Orbital elements in \citet{finger2006} are employed.}\\
      \multicolumn{5}{l}{$^b$:Second peak of the outburst D
        on MJD 55050.}\\
      \multicolumn{5}{l}{$^c$:Second precursor of the outburst E
        on MJD 55180.}\\
    \end{tabular}
  \end{center}
\end{table*}

\begin{figure}
  \begin{center}
  \FigureFile(80mm,){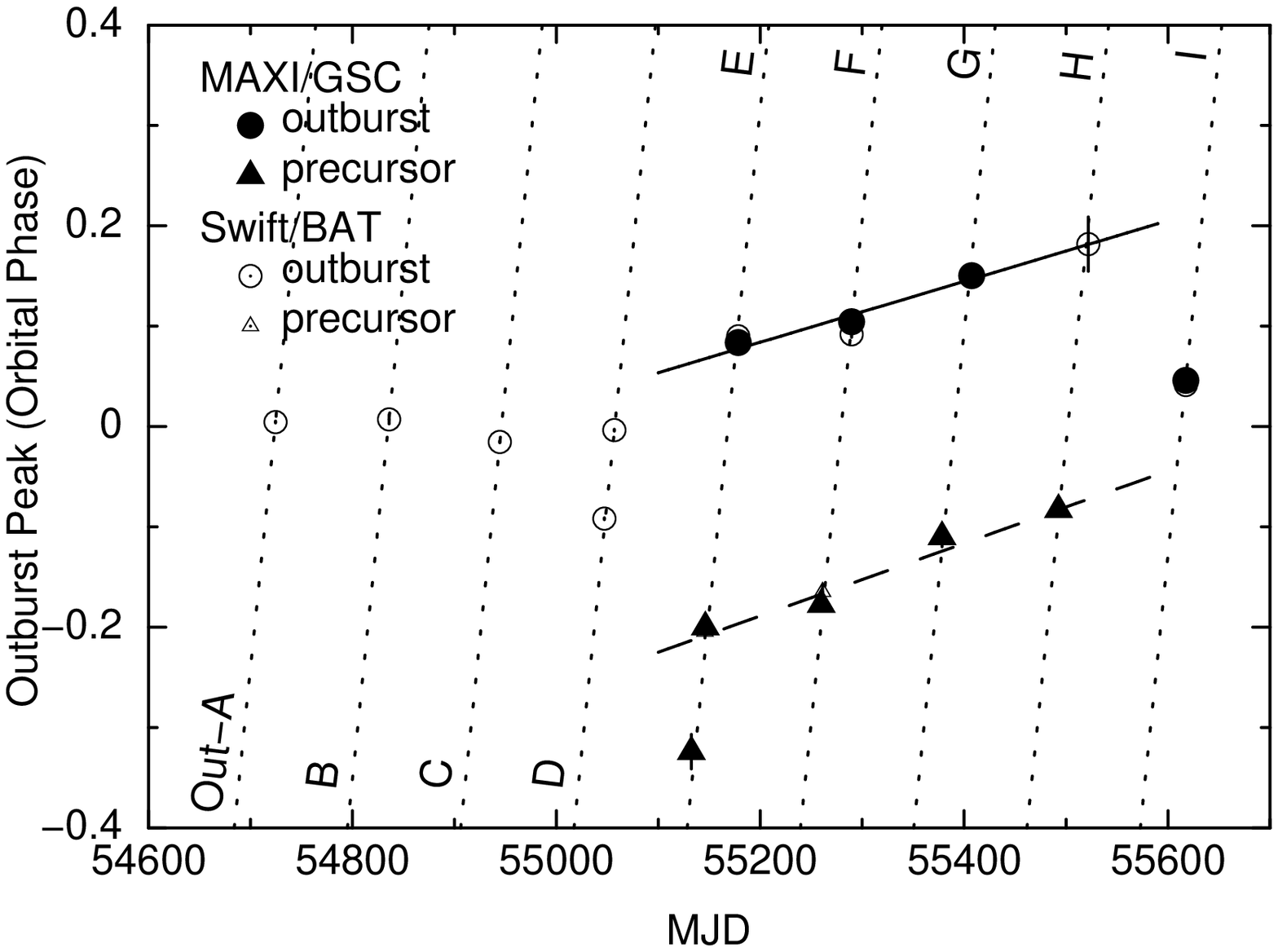}
  \end{center}
  \caption{ Orbital phases of the outburst peaks (circles) and
    precursors (triangles) taken form table \ref{t1}, plotted as a
    function of MJD.  Filled symbols are those obtained from the
    MAXI/GSC (4--10 keV) data, and open ones are from the Swift/BAT
    (15--50 keV) data.  Solid and dashed lines indicate the best-fit
    linear function for the peaks and precursors of Out-E to Out-H,
    respectively.  Dotted lines represent the orbital motion of the
    neutron star. }
\label{f3}
\end{figure}

To be quantitative, we fitted individual outburst / precursor profiles
with a Gaussian function and determined their peak phases.  Since some
outbursts exhibit asymmetric profiles which cannot be reproduced by a
single Gaussian function, we performed the fit within a narrow range
around the peak such that the fit is accepted reasonably.  In
addition, we confirmed that the result does not depend on the chosen
fitting range.

The above analysis has given the best-fit epochs as tabulated in table
\ref{t1}, and plotted in figure \ref{f3}.  The peak phases of Out-A,
B, C and the second peak of Out-D agree with the expected periastron
passage within the precision of 1 d ($=0.009$ in orbital phase).  The
first peak of Out-D emerged $\sim0.1$ orbital phase before the
periastron passage.  After the anomalous Out-D, the four outbursts
(Out-E, F, G and H) show a systematic drift in the phases of the
emission peak and of the precursor, indicating that their recurrent
period is longer than the orbital period.  To evaluate the rates of
the phase shifts, we fitted the large-outburst and precursor data
points in figure \ref{f3} with a linear function, and show the results
by a solid and a dashed lines.  The phase-shift rates of the outbursts
and the precursors were obtained as $(3.03\pm0.06)\times10^{-4}$ phase
day$^{-1}$ and $(3.62\pm0.12)\times10^{-4}$ phase day$^{-1}$,
respectively, with the errors of 1-$\sigma$ confidence limit.  The
phase shift of the precursor is slightly faster than that of the main
outburst.

\begin{figure}
  \begin{center}
    \FigureFile(80mm,){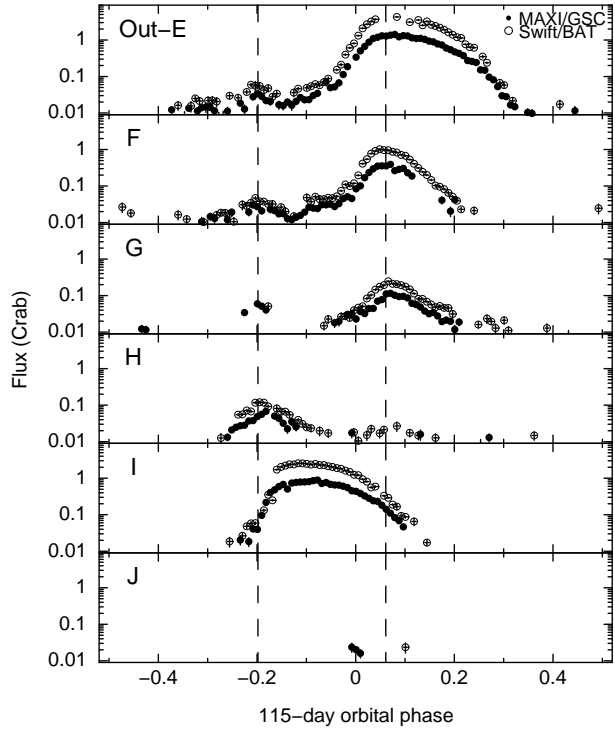}
  \end{center}
  \caption{Same as figure \ref{f2}, but the light curves are folded by
    the best-fit recurrent period of 115 days of the four outburst
    peak from E to H.  Two vertical dashed lines indicate the peak
    positions of the precursors and outbursts in Out-E to H.  The
    115day phase 0 is taken at the true orbital phase 0 of Out-E.  }
\label{f4}
\end{figure}

Next, we evaluate recurrent periods of the outbursts and the
precursors, $P_{\rm out}$ and $P_{\rm pre}$, respectively.  Using the
obtained phase-shift rate $s$ in unit of day day$^{-1}$, $P_{\rm out}$
and $P_{\rm pre}$ are related to the orbital period, $P_{{\rm orb}}$,
as
\begin{equation}
P_{{\rm out, pre}} = \frac{P_{{\rm orb}}}{1-s}.
\label{e1}
\end{equation}
The outburst and precursor periods calculated in this way are
$115.0\pm0.1$ day and $115.8\pm0.2$ day, respectively.  Figure
\ref{f4} shows the light curves folded by the derived outburst beat
period (115.0 day), where the outburst and precursor peaks are both
observed to align well.

\subsection{Evolution of Outburst Peak-Intensity}
\label{sec3.3}

\begin{figure}
  \begin{center}
  \FigureFile(80mm,){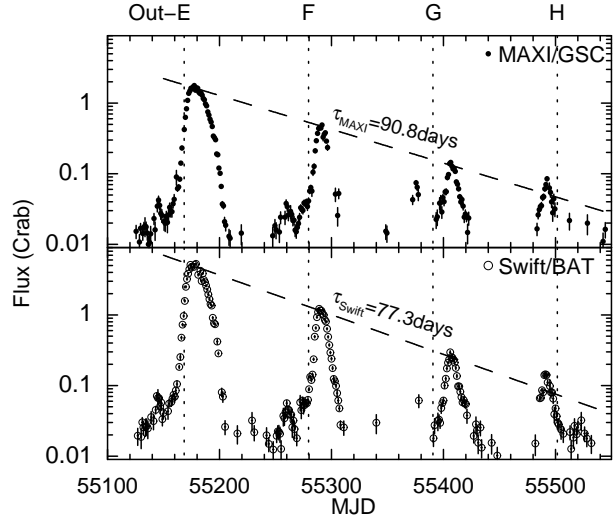}
  \end{center}
  \caption{ The expansion of the four outbursts.  The dotted lines
    represent the periastron passage.  The peak intensities of the
    three outbursts decays exponentially (see text). }
\label{f5}
\end{figure}

As seen in the light curves of figure \ref{f1}, the four outbursts
(Out-E to H) changed not only in their orbital phase, but also in
their intensity.  As shown in figure \ref{f5}, their peak fluxes
decreased exponentially, with an e-folding time of $90.8\pm1.3$ days
estimated with the MAXI/GSC data and $77.3\pm1.4$ days with that of
Swift/BAT.  Here again the errors are in 1-$\sigma$.

After the outburst peak intensities thus decayed exponentially till
Out-H, the next one, Out-I, violated this systematic behavior, and
actually developed into the second largest one.  The next expected
outburst was during the Sun avoidance.  Nevertheless, we can estimate
that the decay time scale should be $<57$ days, considering the lack
of significant signals at two orbital periods later
(MJD$\simeq$55830).

The precursor peak intensities exhibited no systematic changes.
Specifically, the precursors of Out-E and Out-F have almost the same
intensity of $\sim 50$ mCrab (4-10 keV), that of Out-G is not clear
but is probably $\sim 80$ mCrab, and that of Out-H is 100 mCrab.

\section{Discussion}
\label{sec4}

We analyzed long-term X-ray light curves of A 0535+26 obtained by the
MAXI/GSC (4--10 keV) and the Swift/BAT (15--50 keV) during the active
season from 2008 September to 2011 November.  This is the longest
active period of this X-ray source that has ever been observed since
its discovery in 1975.  The source exhibited nine outbursts which are
approximately synchronized with the binary period of 111.1 d.  Four of
them recurred every 115 d which is significantly longer than the
binary period.  In both data sets, we also found low-level active
periods, or precursors, which preceded the large outbursts.  Below, we
discuss the cause of the asynchronous outburst occurrence and of the
precursor activity.

\subsection{Classification of Outbursts -- Normal (Type-I) or Giant (Type-II)}
\label{sec4.1}

Outbursts in Be/X-ray binaries are classified into normal (type-I) and
giant (type-II) ones, according to their luminosity and orbital phase
at the flux peak (e.g. review by \cite{reig2011}).  Normal outbursts
have peak luminosities below $\sim 10^{37}$ erg s$^{-1}$, and occurs
at or near the periastron passage of the neutron star.  Giant
outbursts exhibit higher luminosities exceeding $\sim 10^{37}$ erg
s$^{-1}$, and their orbital phases do not necessarily coincide with
the periastron.  According to the classical definition, we can
classify Out-A, B, and C into normal, while Out-E and I readily into
giant ones.  On the other hand, the other outbursts, Out-D, F, G and
H, cannot be categorized into either type.  As pointed out by
\citet{kretschmar2013}, the simple classification cannot be always
applied.  From the monotonical peak-phase shift (figure \ref{f2} and
figure \ref{f3}) and exponential decay of the peak fluxes (figure
\ref{f5}), we presume that the giant outburst (Out-E) and the
following ones (Out-F, G, and H) have the same mechanism.  Those
followings might be named as "giant outburst remnant".

In addition to the variety in the luminosity and the orbital phase as
described above, Out-D, has a peculiar double-peaked profile.  So far,
outbursts with such a double-peaked profile have been observed three
times from this source, in 1993 (July and November ;
\cite{finger1996}) and 2005 August
\citep{caballero2008,postnov2008,caballero2011,caballero2013}.
However, the latter event is a ``pre-outburst flare'' which lasted for
less than a few hours \citep{postnov2008,caballero2011}; we consider
it as a different phenomenon from the double-peaked outburst such as
Out-D, whose duration ($\sim$10 days) is much longer.  In addition to
A 0535+26, such double-peaked outbursts have been observed from two
other Be/X-ray binaries, XTE J1946+274 \citep{muller2012} and GX304-1
\citep{nakajima2012}.  Among those three sources, A 0535+26 and
GX304-1 exhibited giant outbursts after the detection of a
double-peaked outburst.  Therefore, double-peaked outbursts might be
considered as a sign of the recurrence of giant outbursts
\citep{nakajima2012,caballero2013}, and may be recognized as an
intermediate class between the normal and giant ones.

\subsection{Phase Shift and Intensity Evolution of Giant Outbursts}
\label{sec4.2}

As analyzed in section \ref{sec3}, the outburst-peak phase shifted
steadily through the four giant outbursts, from Out-E to Out-H, and
their peak intensity decayed on a time scale of 77--91 d.  Such an
effect has been observed from two other Be/X-ray binaries : EXO
2030+375 \citep{wilson2002} and GS~0834-430 \citep{wilson1997}.
However, the precursor phase shift observed in the present study was
not reported from the other sources.  Furthermore, the phase shift in
A 0535+26 ($\sim$20 days) is much larger than that of EXO 2030+375
($\sim$5 days).

Since X-ray intensity is considered to provide a probe of the
circumstellar disc of the Be star along the neutron-star orbit, the
outburst phase shift is considered to reflect density profiles in the
Be disc.  Let us assume that dense parts are produced in the disc by
some perturbations, and that they rotates with a certain rotational
frequency, $\nu_{{\rm per}}$.  We further assume that it couples with
the orbital frequency $\nu_{{\rm orb}} = P_{{\rm orb}}^{-1} = 1/111.1$
d$^{-1}$ to produce a beat frequency $\nu_{{\rm beat}}$ as,
\begin{equation}
\nu_{{\rm orb}} - \nu_{{\rm per}} = \nu_{{\rm beat}}, 
\label{e2}
\end{equation}
and identify $\nu_{{\rm beat}}^{-1}$ with the period $P_{{\rm out}} =
115$ d with which the consecutive giant outbursts emerged.  Equation
(\ref{e2}) then yields $\nu_{{\rm per}}=3.13\times10^{-4}$ d$^{-1}$,
which corresponds to a period of $\sim8.7$ years.  Below, we discuss
two possible scenarios to explain the orbital phase shift of the
outbursts including this $\nu_{{\rm beat}}$, referring to figure
\ref{f6} which summarizes the present observations.

\subsection{Global One-armed Oscillation Scenario}
\label{sec4.3}

One possible explanation of the evolution of the outburst profile is a
model of density perturbation called ``global one-armed oscillation''
on a Be disc (e.g. \cite{okazaki1991,papaloizou1992,okazaki1997}),
proposed to explain long-term variations of double-peaked (violet and
red) H$\alpha$ emission lines (\cite{silaj2010} and references
therein).  The violet-to-red intensity ratio, so called V/R ratio,
varies on timescales of years to decades.  The one-armed oscillation
model explains it as due to a slow rotation of a dense part on the
disc.  Although various oscillation modes can be excited when the disc
is formed, those with $m \neq 1$ ($m$ being the azimuthal wave number)
are supposed to dissipate sooner and finally only the one-armed ($m =
1$) oscillation will remain.

\begin{figure}
  \begin{center}
  \FigureFile(85mm,){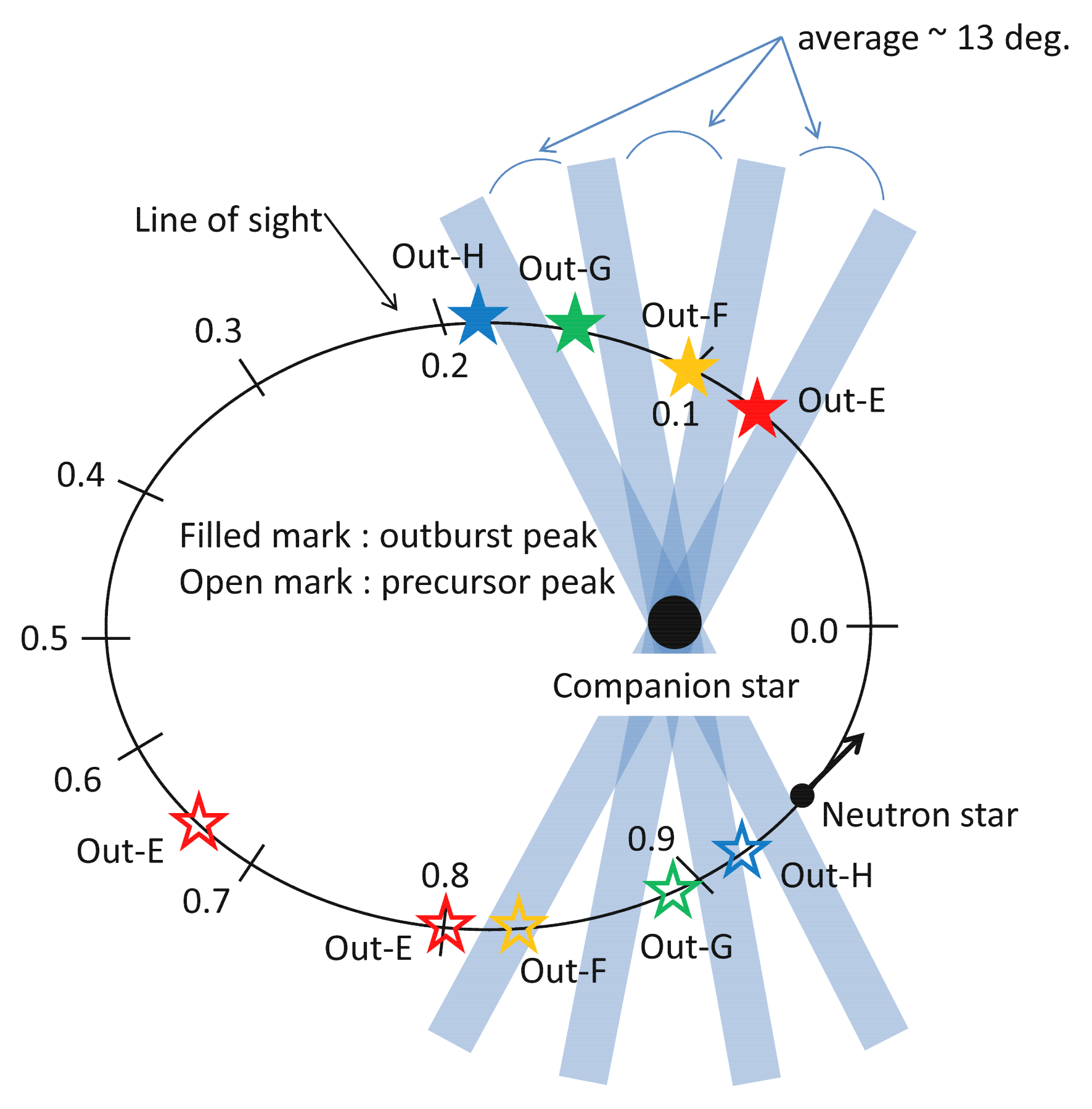}
  \end{center}
  \caption{Orbital geometry of A 0535+26. The open and filled symbols
    represent the positions of the precursor peaks and of the main
    outbursts respectively. }
\label{f6}
\end{figure}

Previous optical studies on A 0535+26 \citep{moritani2010} reported
that the source exhibited V/R variations with a $\sim$500 d
quasi-periodicity from MJD$\sim53700$ to MJD$\sim55000$. It suggests
the emergence of the one-armed oscillation mode.  In that period,
H$\alpha$ EW representing the density at the outer part of the Be disc
and its size, began to increase gradually at around MJD$\sim 54600$
\citep{yan2012,camero2012}.  When the Be disc extended and reached the
periastron of the neutron star, normal outbursts would occur.  This
can explain the three successive normal outbursts from Out-A to Out-C.

While the H$\alpha$ EW increased, the V-band brightness decreased
\citep{yan2012,camero2012}.  This suggests the production of a
low-density region at an inner part of the Be disc \citep{yan2012},
which would expand slowly outwards to produce a low-density ring
structure.  This can account for the double-peak outburst in Out-D
\citep{yan2012}.

After Out-D, the V/R variation did not follow the 500 d periodicity
any longer \citep{moritani2011,moritani2013}, presumably due to some
changes in the physical condition of the Be disc
\citep{negueruela1998}.  In fact, the V-band brightening around MJD
55100 \citep{yan2012} and the very large H$\alpha$ EW around MJD 55150
\citep{yan2012,camero2012} suggest an episodic mass ejection from the
Be-star equator.  The emerged high-density region would propagate to
the outer part of the disc on a time scale of one orbital cycle
\citep{okazaki1991,moritani2011}, and then expand the Be disc
extremely.  The high-density region of the perturbed disc is expected
to rotate slowly around the Be star.  If its period is $\sim8.7\, {\rm
  yr} = \nu_{{\rm beat}}^{-1}$, the observed peak-phase shift of the
outbursts can be explained.

Besides the outbursts, the precursors also emerged with 115 d
periodicity from Out-E to Out-H.  Since the precursors appeared at
nearly opposite orbital locations of the giant outbursts (figure
\ref{f6}), they may be explained by the density perturbation in an
unstable $m=2$ oscillation mode which embeds a two-arm structure.

The estimated rotational period of the density perturbation ($\sim8.7$
yr) is considerably longer than the typical V/R variation period of
$1\sim1.5$ yr \citep{clark1998,moritani2010} and comparable to the
statistical mean of the V/R variation periods of isolated Be stars,
$\sim7$ yr (\cite{okazaki1991} and references therein).  Optical
observations of A 0535+26 (figure 2 of \cite{moritani2011} and figure
4 of \cite{camero2012}) indicate that there would be a longer V/R
variation period when the outburst orbital phase shift appeared.  This
anomalous V/R variation lasted for $\sim1$ yr, and vanished when a
rapid variation (period of $\sim25$ d) arose in 2010 October
\citep{camero2012}.

\subsection{Be-disc Precession Scenario}
\label{sec4.4}

\begin{figure}
  \begin{center}
  \FigureFile(80mm,){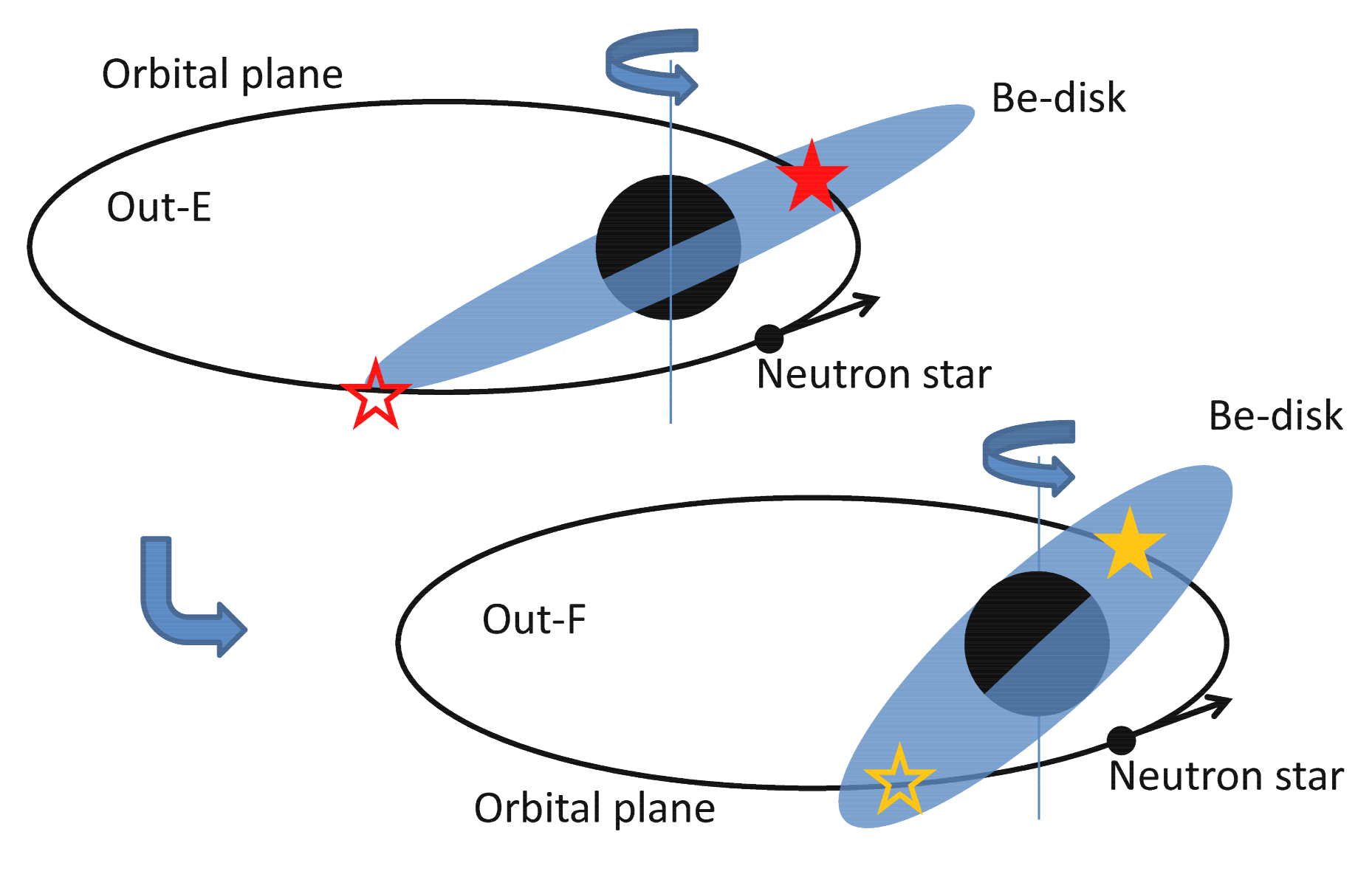}
  \end{center}
  \caption{ Schematic illustration of the Be-disc precession scenario,
    in which precursors and giant outbursts occur at the moving
    intersections between the neutron star orbit and the Be disc.  }
  \label{f7}
\end{figure}

Another scenario to explain the orbital phase shift of the outbursts
is a Be-disc precession.  The idea is proposed by
\citet{negueruela2001} to explain the behaviors of 4U 0115+63.
According to this scenario, a Be disc does not grow as an equatorial
planar disc any more when a large amount of gas is emitted.  Instead,
the disc is considered to be warped by the interaction with the
neutron star, and tilted.  The tilted disc starts precessing.  The
scenario, applied to A 0535+26, is illustrated in figure \ref{f7}.
\citet{okazaki2013} and \citet{moritani2013} also examined a similar
tilted Be-disc geometry to explain the variety of observed outburst
behaviors, which are largely classified into the normal or the giant
one.  They suggested that giant outbursts could occur if a Be disc is
misaligned with a binary orbital plane and sufficiently developed so
that a neutron star can capture a large amount of mass via
Bondi-Hoyle-Lyttleton (BHL) accretion.  We here employ the tilted
Be-disc geometry to explain the observed outburst orbital-phase
shifting.

Both the optical and X-ray observations suggest that a large amount of
gas was ejected from the Be-star equator after Out-D as described in
subsection \ref{sec4.3}.  As a consequence, a tilted
and sufficiently developed disc may be formed and
start a prograde precession.  In fact, we can see in figure \ref{f6}
that each outburst and the associated precursor occurred at nearly
opposite orbital locations with respect to the Be star, and their peak
locations were delayed every binary orbit by $0.02\sim0.04$ orbital
phase.  These results can be explained by such a scenario that the
precursor and the main outburst occur at the two intersections between
the neutron star orbit and the Be disc plane.  The recurrence period
of the giant outbursts, 115 d, can be explained if the disc precesses
by the phase shift rate $s$, which is $13^\circ$ per pulsar's orbital
period of 111.1 d (see figure \ref{f6}).  We also noticed that there
is a inconsistency in the Be disc precession period between our result
($\sim8.7$ yr) and the previous ones (674 or 886 d ;
\cite{moritani2013}), although such a comparison is beyond the scope
of this paper.  The intensity of precursors grew up with time, since
these orbital phase became closer to the periastron where the disc
density is thickest.  The main peak intensity declined vice verse as
it became further away from the Be star with time.

In this scenario, the Out-H peak at around MJD$=55490$ is considered
as a precursor, and the smaller peak later at around MJD$=55520$ is
considered as a main outburst.  Thus, the intensities of the main
peaks and the precursors of Out-E, F, G, and H can be considered to
represent the density change with time of the tilted Be disc formed at
around Out-D.  This scenario thus gives a reasonable explanation to
the present observation, except the first of the two precursors of
Out-E : another smaller tilted or warped disc would be required for
it.  Further modeling of the Be disc is left for future works.

\section{Conclusion}
\label{sec5}

Over the 2009 August -- 2011 March active period of the Be/X-ray
binary pulsar A 0535+26, nine outbursts were detected, of which two
are categorized into giant (type-II) outbursts according to the peak
intensity and the orbital phase.  The giant outbursts and subsequent
outbursts recurred with a period of 115 d, which is significantly
longer than the 111.1 d orbital period.  Prior to these phase-shifting
outbursts, we detected the precursors, which also repeated with the
115 d period.  This period shift is considered to reflect some
structure, of the circumstellar disc formed around the Be companion
star.  If such structures have an intrinsic period of $\sim8.7$ yr,
the observed giant-outburst period can be explained by its beat with
the orbital period.  Two possible scenarios for such a disc
perturbation were discussed.  One is the one-armed density
perturbation on the Be disc, which is inferred from the optical
spectroscopy of H$\alpha$/He$_{\rm I}$ emission lines.  The other is
the precession of the Be disc, which is suggested by the orbital
positions of the giant outbursts and precursors.  The latter scenario
reminds us a model of giant outbursts proposed by Okazaki et
al. (2013) that they could occur when a misaligned Be disc is
developed sufficiently.  Further modeling of the Be disc is left for
future works.

\bigskip

We appreciate all MAXI team members for their dedicated works to
enable the science analysis.
This research has made use of Swift/BAT transient monitor results
provided by the Swift/BAT team at the Goddard Space Flight Center and
Los Alamos National Laboratory.
The present research work is partially supported by the Ministry of
Education, Culture, Sports, Science and Technology (MEXT),
Grant-in-Aid for Science Research 24340041.


\end{document}